\documentclass[prd,aps,amsmath,amssymb,showpacs,11pt]{revtex4-1}

\usepackage{amssymb}
\usepackage{amsbsy}
\usepackage{bm}
\usepackage{latexsym}
\usepackage[active]{srcltx}
\usepackage{latexsym,amsfonts,amsthm,amsmath,amscd,amssymb,color}
\newcommand{\sect}[1]{\setcounter{equation}{0}\section{#1}}

\numberwithin{equation}{section}

\begin{document}

\title{\bigskip\bigskip\bigskip\bigskip\bigskip
	{Relativistic fine structure of the $\bigl({}^4{\mathrm{He}},\mu\bigr)$ ion $2p$-levels}}

\author{{R. Giachetti}}
\address{Physics Department, Universit\`a di Firenze, Italy}
\author{{E. Sorace}}
\address{{Istituto Nazionale di Fisica Nucleare, Sezione di Firenze, Italy}}

\begin{abstract}
Motivated by the recent experimental measures of the $2p$-states fine 
splitting of the muonic ${\mathrm{He}}^4$ ion \cite{Nat}, we think it 
may have some interest to present an 
independent calculation in terms of the two-body 
relativistic equation for a scalar and a fermion developed in 
\cite{GSM,GSR} for mesic atoms. This work can thus be considered an 
addendum to these last papers in a different range for the masses  of
the two components. It gives theoretical results in good agreement 
with the recent experimental measures.  

 
\end{abstract}
\maketitle

\voffset= 1.0 truecm

\sect{Introduction.} \label{Sec_intro}
\bigskip

The data of very accurate experimental measures of the  fine splitting 
of the muonic $\mathrm{He}^4$ ion have been recently published in
\cite{Nat}. The  reported energies for the 
$2p_{3/2}\rightarrow 2s_{1/2}$ and 
$ 2p_{1/2}\rightarrow 2s_{1/2}$ transitions are respectively 
$368653\pm 18$  GHz = $1524.6258\pm 0.0744$ meV and 
$333339\pm 15$ GHz = $1378.5789\pm 0.6534$ meV.
The experimental fine splitting $ 2p_{3/2}\rightarrow 2p_{1/2}$
is finally  found to be $146.047\pm 0.096$ meV.
In the same paper the theoretical results, mainly referring to 
\cite{Diep,KKSI,MartEl,Ji,Borie}, have also been presented. 
The slightly different
theoretical $2p_{1/2}\rightarrow 2s_{1/2}$ splittings 
obtained by different groups are reported in \cite{Diep} where the
chosen value is $1668.4892\pm 0.0135$. This figure is independent of 
the nuclear structure and contains  the non-relativistic one- and 
two-loop electron vacuum polarization and the corresponding 
relativistic corrections.  
The nuclear contributions are estimated to be 
$9.340\pm 0.250$ meV + $0.0112$ meV. 
The mismatch between theoretical and experimental data of the 
$2p_{1/2}\rightarrow 2s_{1/2}$ transition is overcome by taking
into account the charge radius $r_{\mathrm{nucl}}$ of 
the ${\mathrm{He}^4}$ nucleus.  As shown in \cite{Diep,KKSI,MartEl,Ji,Borie} it is expressed by the relation 
$\Delta E = A r_{\mathrm{nucl}}^2 + B$. 
Although large differences may appear 
in the parameters $A$ and $B$ presented by different research groups, 
the final values of the ${\mathrm{He}^4}$ nucleus charge radius
are however  very close. In  \cite{Diep}, for instance, the choice is
$A=-106.3536~{\mathrm{meV}}{\mathrm{/fm}}^2$ and
$B=0.0784~{\mathrm{meV}}$, while in
\cite{Nat} the authors assume $A=-106.220~{\rm{meV}}{\rm{/fm}}^2$,
$B=0.00112~{\mathrm{meV}}$. Since the presented 
$2p_{1/2}\rightarrow 2s_{1/2}$ splittings are $1677.6792~{\rm{meV}}$
and $1677.670~{\mathrm{meV}}$ respectively, the corresponding results 
for $r_{\mathrm{nucl}}$ are $1.677~\mathrm{fm{}}$ and 
$1.678~\mathrm{fm{}}$. 

The usual and well established method to obtain
theoretical results is to start from a non-relativistic atomic 
framework and calculate the corrections due to relativity,
to the two-body recoil, to QED electron vacuum polarization
and the contribution of nuclear properties, mainly 
involving the finite charge radius of the  
nucleus. When the light particle is a fermion, the Dirac equation has 
also been used for a better initial description of the system 
\cite{Borie}. In last years we have proposed a scheme
allowing the formulation of relativistic two-body wave equations, 
according to their fermionic or bosonic nature. Details and 
references to previous papers can be
found in \cite{GSM,GSR}. Results in good
agreement with experimental data have been produced
through a wide range of energies, from mesons to atoms. In
particular, in \cite{GSM} we have dealt with mesic atoms
constituted by a proton and a scalar meson,  $\pi$ or  $K$. 
This last framework applies also to the
$\bigl({}^4{\mathrm{He}},\mu\bigr)$ ion, with the difference that
the scalar is now the heavy particle. We thus believe that there 
may be some interest in producing our results, obtained in an
independent covariant, two-body scheme and to compare them with 
the present experimental data. 
  
\sect{Statement of the results.} \label{Sec_results}

We solve the spectral problem for the lowest levels of the
relativistic scalar-fermion $\bigl(\,{}^4{\mathrm{He}},\mu\,\bigr)$
ion using the two-body equation given in \cite{GSM}. The physical
parameters we use are the following. The Helion mass
is $m_{\mathrm{He}} = 3727.3794066$ MeV, the muon mass is  
$m_{\mathrm{\mu}} =105.6583755$ MeV, the fine
structure constant is $\alpha = 0.0072973525693$ \cite{NIST}.
The results for the lowest levels are reported 
in the following Table I.
\begin{table}[h]
	\begin{center}
		\begin{small}
			{{ \begin{tabular}{|c||r|r|r|r|}
						\hline 
						State
						&$1s_{1/2}$\phantom{9011}
						&$2s_{1/2}$\phantom{9011}
						&$2p_{1/2}$\phantom{9011}
						&$2p_{3/2}$\phantom{9011}
					\\
						\hline 
						Levels
						& -10943.19011297
						&  -2735.83861160
						&  -2735.84873222
						&  -2735.71095812
						
						\\
						\hline
					\end{tabular}
			}}
		\end{small}
		\caption{The lowest pure Coulomb levels of the $\bigl({}^4{\mathrm{He}},\mu\bigr)$ ion in eV.}
	\end{center}
\end{table}

Observe that the the $2s_{1/2}$ and the $2p_{1/2}$ are not degenerate
as for a Dirac electron, since the Johnson-Lippman symmetry 
\cite{GSM} is broken in the two-body equation. 
We then determine the electron vacuum polarization (eVP) corrections 
of the levels by calculating the corresponding
 matrix elements of the  Uehling, K\"allen-Sabry and 
iterated reducible Uehling potential with the relativistic 
eigenfunctions. The use of the latter removes the  requirement of
further relativistic corrections on the result.
The expressions of the potentials have been taken 
from \cite{Ji,KKI}. Their values are given in Table II.

\begin{table}[h]
	\begin{center}
		\begin{small}
			{{ \begin{tabular}{|c||r|r|r|r|}
						\hline 
						State
						&$1s_{1/2}$\phantom{9011}
						&$2s_{1/2}$\phantom{9011}
						&$2p_{1/2}$\phantom{9011}
						&$2p_{3/2}$\phantom{9011}
						\\
						\hline 
						Uehling
						&-18792.31849 
						& -2077.94106  
						&  -411.74833  
						&  -411.48797  
							\\
						\hline 
						 K\"allen-Sabry
						&  -108.96659 
						&   -11.77264  
						&    -3.88856  
						    &-3.88671  
							\\
						\hline 
						Two loop reducible
						&   -30.37895 
						&    -3.64576  
						&     0.15541  
						&     0.15561  
						\\
						\hline
					\end{tabular}
			}}
		\end{small}
		\caption{The eVP corrections to the lowest levels of the $\bigl({}^4{\mathrm{He}},\mu\bigr)$ ion in meV.}
	\end{center}
\end{table}

Thus the total eVP correction to the 
$2p_{1/2}\rightarrow 2s_{1/2}$ splitting is $1677.8780$ meV  and 
to the $2p_{3/2}\rightarrow 2p_{1/2}$ splitting is $0.26241$ meV.
Including the above QED corrections we therefore have
$$\Delta_{2p_{1/2}\rightarrow 2s_{1/2}} = 1667.7574~{\mathrm{meV}}$$
This has to be compared with the value $1668.489(14)$ meV, reported in
\cite{Nat}. These last two figures must be corrected with nuclear
properties. As these are not pure QED effects, we will assume the same 
formula (11) of \cite{Nat} in order to evaluate the contributions and
consequently we find a Helion charge radius equal to $1.676$ fm.

Let us turn to the fine structure of the $p$-states. The value of
the $2p_{3/2}\rightarrow 2p_{1/2}$ splitting given in \cite{Nat} and
taken from \cite{Diep}, is determined by the the Dirac splitting of 
the $p$-levels to which four different groups of corrections are
added. The first group collects the contributions due to relativity
and recoil. Adding these, the proposed splitting is $145.5833$ meV. 
The second group considers the eVP with the corresponding
relativistic corrections and the third
one takes into account the $\mu$ anomalous magnetic moment. The fourth
and last group deals with the finite
charge radius. The final figure proposed for the
$2p_{3/2}\rightarrow 2p_{1/2}$ splitting is $148.1828$ meV.

We give a short discussion of our results. The previous four groups 
of contributions can be distinguished in our framework also.
The first three of them, however, come from different inputs as we  
later explain. The correction for the finite
nuclear charge is not included in our scheme and we assume for it
the same value proposed in \cite{Diep}, namely $\,-0.0113$ meV.
The corrections for the muon anomalous magnetic moment and the one for 
the `external' part of the fine structure are neither included: 
indeed, while the spin-orbit of the muon with itself is taken into
account in the two body relativistic equation \cite{GSR,GSM},  
the interaction between the spin of the muon and the Helion nucleus
orbit is not included. Its contribution is therefore determined by 
means of the Pauli-Breit perturbative term, as generally done 
\cite{BS,MartEl}. The anomalous magnetic moment is taken into account 
in a similar way. The QED corrections for the eVP polarization,
calculated using the two-body relativistic wave-functions, are
finally added. 

The fine splitting of the $2p$-states including the difference of 
the $p$-levels $\Delta^{0}_{2p_{3/2}\rightarrow 2p_{1/2}}=137.7741$ 
meV, purely quantum mechanical and thus non-perturbative, 
together with the muon external 
spin-orbit and its anomalous magnetic moment $a_{\mu}=0.001165920$, is
$$ \Delta^{\mathrm{QM}} = \Delta^{0}_{2p_{1/2}\rightarrow 2s_{1/2}}~
\biggl(\,1 + 2\frac{m_{\mu}}{m_{\mathrm He}} 
+ 2\,a_{\mu}\Bigl(1 + \frac{m_{\mu}}{m_{\mathrm He}}\Bigr)\,\biggr)
= 145.9153~{\mathrm{meV}}$$
The level splitting $\Delta^0$ is given in literature 
by the approximate perturbative  expression 
$(Z\alpha)^4\,m_R^3/(32\,m_{\mu}^2) = 137.75798$ meV \cite{MartEl}. The
eVP corrections are obtained from Table II and amount to
$$\Delta^{\mathrm{QED}} = 0.2624053133 ~{\mathrm{meV}}$$
Assuming finally the value $\Delta^{\mathrm{N}}=-0.01176$ meV for
the finite nuclear  charge radius correction \cite{Diep}, our
final result for the $2p$ fine structure of the 
$\bigl(\,{}^4{\mathrm{He}},\mu\,\bigr)$ ion is
$$\Delta_{2p_{3/2}\rightarrow 2p_{1/2}} = 146.1660~{\mathrm{meV}}$$
in good agreement with experimental measures.

\bigskip


\end{document}